\documentclass[12pt]{article}

\begin{document}

\input amssym.tex

\title{Relativistic currents on  ideal Aharonov-Bohm cylinders}

\author{Ion I. Cot\u aescu, Doru-Marcel B\u alt\u a\c teanu and Ion  Cot\u{a}escu Jr. \\
{\small \it West University of Timi\c{s}oara,}\\
{\small \it V.  P\^{a}rvan Ave.  4, RO-300223 Timi\c{s}oara, Romania}}

\maketitle

\begin{abstract}
The relativistic theory of the Dirac fermions moving on cylinders in external Aharonov-Bohm field  is built starting with a suitably restricted Dirac equation whose spin degrees of freedom are not affected. The exact solutions of this equation on finite or infinite  Aharonov-Bohm cylinders allow one to derive the relativistic circular and longitudinal currents pointing out their principal features. It is shown that all the circular currents are related to the energy in the same manner on cylinders or rings either in the relativistic approach or in the non-relativistic one. The specific relativistic effect is the saturation of the circular currents for high values of the total angular momentum. Based on this property some approximative closed formulas are deduced for the total persistent current at $T=0$ on finite Aharonov-Bohm cylinders. Moreover, it is shown that all the persistent currents on finite cylinders or rings have similar non-relativistic limits.    
\end{abstract}


\newpage

\section{Introduction}

The non-relativistic quantum mechanics based on the Schr\" odinger  equation is the principal framework for studying the electronic effects in mesoscopic systems  \cite{B1}-\cite{B10} were  the spin-orbit interaction is described by additional terms \cite{B11}-\cite{B15}. 
However, there are  nano-systems, as for example the graphenes,  where several relativistic effects observed in the electronic transport  can be satisfactory explained  considering the electrons as  massless Dirac particles moving on honeycomb lattices \cite{N1}-\cite{Y2}.  

For this reason,  many  studies  \cite{N1}-\cite{B20} focus on the relativistic effects considering the electrons near the Fermi surface as being described by  the $(1+2)$-dimensional Dirac equation corresponding to a restricted three-dimensional Clifford algebra. This method  restricts simultaneously not only the orbital degrees of freedom but the spin ones too, reducing them to those of the $SO(1,2)$ symmetry and affecting thus the spin-orbit coupling. Starting with this remark, we have shown that  there are situations when it is convenient to use the {\em complete} $(1+3)$-dimensional Dirac equation,  restricting the orbital motion, according to the concrete geometry of the studied system, but without affecting the natural spin degrees of freedom given by the $SL(2,{\Bbb C})$ symmetry. Thus the spin-orbit coupling remains unharmed describing  the orbital and spin effects with more accuracy. 

A special problem that was treated only occasionally with the complete  Dirac equation  is that of the fermions in external Aharonov-Bohm (AB) field  \cite{scat1} - \cite{MIT}.   The persistent currents in AB rings were recently studied starting with  a version of the  Dirac equation in which the orbital restrictions induce an unwanted non-Hermitian term  \cite{CP} that  distorts the results,  even though these may outline new realistic effects  \cite{Ghosh}. For this reason, we proposed a new approach based on  the {\em correctly} restricted Dirac equation, involving only Hermitian terms,  that can be obtained easily starting with a suitable restricted Lagrangian theory \cite{CBC}. We derived thus the expression of the relativistic currents in AB rings pointing out the saturation effect and  deducing the form of the corresponding  persistent currents at $T=0$ \cite{CBC}.

In the present paper we would like to extend this study to the relativistic ideal AB cylinders considered as geometric manifolds without internal structure.  We derive the fundamental solutions of the  Dirac equation on  AB cylinders as common eigenspinors of a  complete systems of commuting operators including the energy, total angular momentum and a specific operator analogous to the well-known Dirac spherical operator of the relativistic central problems \cite{TH}. Appropriate boundary  conditions define different  solutions, on infinite or finite AB cylinders, that can be normalized with respect to the relativistic scalar product. We obtain thus the systems of normalized fundamental solutions that allow us to write down the exact expressions of the relativistic circular and longitudinal currents and to derive the persistent circular currents on finite AB cylinders at $T=0$.

First we find that the circular currents on finite or infinite AB cylinders remain related to the derivative of the relativistic energies with respect to the flux parameter just as in the cases of the relativistic or even non-relativistic AB  rings \cite{CBC}. However, the principal relativistic effect is the saturation of the circular currents on finite or infinite AB cylinders in the limit of very high total angular momenta.  It is remarkable that this effect is the same on AB cylinders or rings, the saturation value depending only on the radius of their transverse sections \cite{CBC}. 

The paper is organized as follows. In the second section we present the relativistic theory of the fermions in AB cylinders based on a suitable restriction of  the complete Dirac equation. The fundamental solutions on infinite AB cylinders are derived in the third section pointing out the saturation effect. Ne next section is devoted to the solutions on finite AB cylinders determined by boundary conditions of MIT type that eliminate the longitudinal current but preserving the saturation properties of the circular one. In the fifth section we derive the persistent circular currents at $T=0$ on finite AB cylinders discussing the case of very short ones and the non-relativistic limit.  Finally we briefly present our conclusions.

\section{Dirac fermions on AB cylinders}

We consider the motion of a Dirac fermion of mass $M$  on an {\em ideal} cylinder of radius $R$  whose axis is oriented along the homogeneous and static external magnetic field $\vec{B}$ given by the electromagnetic potentials $A_0=0$ and $\vec{A}=\frac{1}{2}\vec{B}\land \vec{x}$.  This background  is a two-dimensional manifold (without internal structure) embedded in the three-dimensional space obeying the simple equation  $r=R$  in cylindrical coordinates $(t,\vec{x}) \to (t, r, \phi, z)$ with the $z$ axis oriented along $\vec{B}$.

Then, it is natural to assume that any field $\psi$ defined on this manifold depends only on the remaining coordinates $(t,\phi,z)$ such that we can put  $\partial_r\psi=0$ in the kinetic term of the Lagrangian density. Thus we obtain the action of the Dirac fermion in the mentioned external magnetic field
\begin{equation}
{\cal S}={\cal S}_0-\beta\int dt\, d\phi\, dz\,\overline{\psi}\gamma^{\phi}\psi
\end{equation}
having the kinetic part  
\begin{eqnarray}
{\cal S}_0&=&\int dt\, d\phi \,dz\,\left\{\frac{i}{2}\left[\overline{\psi}(\gamma^0\partial_t\psi +\gamma^{\phi}\partial_{\phi}\psi+\gamma^3\partial_z\psi)\right.\right.\nonumber\\ &&-\left.\left.(\partial_t\overline{\psi}\gamma^0+\partial_{\phi}\overline{\psi}\gamma^{\phi}+\partial_z\overline{\psi}\gamma^3)\psi \right]-M\overline{\psi}\psi\right\}\,,
\end{eqnarray}
where  $\overline{\psi}=\psi^{\dagger}\gamma^0$ and 
\begin{equation}\label{gamph}
\gamma^{\phi}=\frac{1}{R}(-\gamma^1\sin\phi+\gamma^2\cos\phi)\,.
\end{equation}
The notation $\beta=\frac{1}{2}eBR^2$ stands for the usual dimensionless flux parameter (in natural units).

From this action we obtain  the correctly restricted Dirac equation, $E_D\psi=M\psi$,  with  the  self-adjoint Dirac operator   
\begin{equation}\label{D2}
E_D=i\gamma^0\partial_t +\gamma^{\phi}(i\partial_{\phi}-\beta)+\frac{i}{2}\,\partial_{\phi}(\gamma^{\phi})+i\gamma^3\partial_z\,,
\end{equation}
whose supplemental third term guarantees that $\overline{E}_D=E_D$. 
This operator commutes with the energy operator $H=i\partial_t$, the momentum along to the third axis, $P_3=-i\partial_z$, and the similar component, $J_3=L_3+S_3$, of the total angular momentum, formed by the orbital part $L_3=-i\partial_{\phi}$ the spin one $S_3=\frac{1}{2}\,{\rm diag} (\sigma_3,\sigma_3)$. 

 Under such circumstances, we have the opportunity to look for particular solutions of the form
\begin{equation}\label{psi}
\psi_{E,\lambda}(t, \phi,z)=N\left(
\begin{array}{c}
f_1(z)e^{i\phi(\lambda-\frac{1}{2})}\\
f_2(z)e^{i\phi(\lambda+\frac{1}{2})}\\
g_1(z)e^{i\phi(\lambda-\frac{1}{2})}\\
g_2(z)e^{i\phi(\lambda+\frac{1}{2})}
\end{array}\right) e^{-iEt}
\end{equation}
which satisfy the eigenvalue problems,
\begin{equation}
H\psi_{E,\lambda}(t, \phi,z)=E\psi_{E,\lambda}(t, \phi,z)\,, \quad
J_3\psi_{E,\lambda}(t, \phi,z)=\lambda\psi_{E,\lambda}(t, \phi,z)\,,
\end{equation}
laying out the energy $E$ and  the quantum number $\lambda=\pm\frac{1}{2},\pm \frac{3}{2},...$ of the total angular momentum.
The normalization constant $N$ has to be determined after we solve the functions of $z$ from the remaining reduced equation that  in the standard representation of the gamma matrices (with diagonal $\gamma^0$) reads
\begin{equation}
\left(
\begin{array}{cccc}
E-M &0&i\partial_z &\frac{i}{R}(\lambda+\beta)\\
0&E-M &-\frac{i}{R}(\lambda+\beta)&-i\partial_z\\
-i\partial_z&-\frac{i}{R}(\lambda+\beta)&-E-M&0\\
\frac{i}{R}(\lambda+\beta)&i\partial_z&0&-E-M
\end{array}\right)\,\left(
\begin{array}{c}
f_1(z)\\
f_2(z)\\
g_1(z)\\
g_2(z)
\end{array}\right) =0\,.
\end{equation}
This is in fact a system of linear differential equations allowing us to solve the functions of $z$.

The general solutions of this system can be obtained reducing the number of functions with the help of the last two equations that yield
\begin{equation}\label{g12}
\left(
\begin{array}{c}
g_1(z)\\
g_2(z)
\end{array}\right)=\frac{1}{E+M}\left(
\begin{array}{cc}
-i\partial_z& -\frac{i}{R}(\lambda+\beta)\\
\frac{i}{R}(\lambda+\beta)&i\partial_z
\end{array}\right)\left(
\begin{array}{c}
f_1(z)\\
f_2(z)
\end{array}\right)
\end{equation}
leading to the second order equations
\begin{equation}\label{ene}
\left(E^2-M^2-\frac{1}{R^2}(\lambda+\beta)^2 +\partial_z^2\right)f_{1,2}(z)=0\,.
\end{equation}
Consequently, the solutions must be linear combinations of the form
\begin{equation}
f_{1,2}(z)=c_{1,2} e^{ikz}+c'_{1,2}e^{-ikz}
\end{equation}
 where $k$ is the fermion momentum along the $z$ axis.  The concrete form of these solutions depends on the boundary conditions we chose for determining  the integration constants $c_{1,2}$ and $c'_{1,2}$ up to a normalization factor, $N$.  This last constant has to be determined by imposing the desired  normalization condition with respect to the relativistic scalar product
 \begin{equation}
 \langle \psi, \psi'\rangle=R\int_{0}^{2\pi}d\phi\int_{D_z}dz \psi^{\dagger}(t,\phi,z) \psi'(t,\phi,z)\,,
 \end{equation}
calculated on the domain $D_z$ of the entire cylinder.

\section{Currents on infinite AB cylinders}

The simplest case is of the infinite cylinder, with $D_z={\Bbb R}$,  where the motion along its axis is a free one. We assume that the spin projections are measured just with respect to this axis such that  we may chose
\begin{equation}
\left(
\begin{array}{c}
f_1(z)\\
f_2(z)
\end{array}\right)=Ne^{ikz} \xi_{\sigma}\,, \quad k\in {\Bbb R}\,,
\end{equation}
where $\xi_{\sigma}$ are the usual Pauli spinors, 
\begin{equation}
\xi_{\frac{1}{2}}=\left(
\begin{array}{c}
1\\
0
\end{array}\right)\,,\quad \xi_{-\frac{1}{2}}=\left(
\begin{array}{c}
0\\
1
\end{array}\right)\,.
\end{equation}
 of polarizations $\sigma=\pm\frac{1}{2}$. Then the functions $g_{1,2}(z)$ can be derived from Eq. (\ref{g12}) as
\begin{equation}
\left(
\begin{array}{c}
g_1(z)\\
g_2(z)
\end{array}\right)=\frac{ Ne^{ikz}}{E+M}\left(
\begin{array}{cc}
 k& -\frac{i}{R}(\lambda+\beta)\\
\frac{i}{R}(\lambda+\beta)&- k
\end{array}\right) \xi_{\sigma}\,,
\end{equation}
while the energy that depends on $k$ and $\lambda$ reads
\begin{equation}\label{Ekm}
E_{k,\lambda}=\left[ M^2+k^2+\frac{1}{R^2}\left(\lambda+\beta \right)^2  \right]^{\frac{1}{2}} 
\end{equation}
as it results from Eq. (\ref{ene}). We obtain thus a mixed energy spectrum whose ground level is given by $k=0$ and one of the values $\lambda=\pm\frac{1}{2}$ that minimizes the last term in Eq. (\ref{Ekm}), e. g. $\lambda=-\frac{1}{2}$ if $\beta>0$. Consequently, the spinor components have to be tempered distributions that may be normalized in the momentum scale.

According to the above results we can write  two types of fundamental solutions   (\ref{psi}) of the form 
\begin{equation}
U_{k,\lambda}^{\pm}(t,\phi,z)=u^{\pm}_{k,\lambda}(\phi)\frac{1}{\sqrt{2\pi}}e^{-iE_{k,\lambda}t +ikz}\,,
\end{equation}
corresponding to the polarizations $\sigma =\pm\frac{1}{2}$. For $\sigma =\frac{1}{2}$ we obtain
\begin{equation}\label{psiUp}
u_{k,\lambda}^+(\phi)=N_{k,\lambda}^+\left(
\begin{array}{c}
e^{i\phi(\lambda-\frac{1}{2})}\\
0\\
\frac{k}{E_{k,\lambda}+M}e^{i\phi(\lambda-\frac{1}{2})}\\
\frac{i(\lambda+\beta)}{R(E_{k,\lambda}+M)}e^{i\phi(\lambda+\frac{1}{2})}
\end{array}\right) 
\end{equation}
assuming that the normalization factor depend on $k$ and $\lambda$.
Similarly, for $\sigma=-\frac{1}{2}$ we deduce
\begin{equation}\label{psiUm}
u_{k,\lambda}^-( \phi)=N_{k,\lambda}^-\left(
\begin{array}{c}
0\\
e^{i\phi(\lambda+\frac{1}{2})}\\
\frac{-i(\lambda+\beta)}{R(E_{k,\lambda}+M)}e^{i\phi(\lambda-\frac{1}{2})}\\
\frac{-k}{E_{k,\lambda}+M}e^{i\phi(\lambda+\frac{1}{2})}
\end{array}\right) 
\end{equation}
It remains to calculate the normalization in the momentum scale finding that by fixing the values
\begin{equation}
N_{k,\lambda}^{\pm}=\frac{1}{\sqrt{2\pi R}}\sqrt{\frac{E_{k,\lambda}+M}{2E_{k,\lambda}}}\,,
\end{equation}
we obtain the desired (generalized) orthogonality relations
\begin{equation}
\langle U^{\pm}_{k,\lambda}, U^{\pm}_{k',\lambda'}\rangle=\delta_{\lambda,\lambda'}\delta(k-k')\,,\quad \langle U^{\pm}_{k,\lambda}, U^{\mp}_{k',\lambda'}\rangle=0\,.
\end{equation}

These fundamental solutions are  eigenspinors of the same set of commuting operators which seems to be incomplete as long as we cannot distinguish between $U^+$ and $U^-$. Therefore we need to introduce a new operator for completing this set. A short inspection suggests that this must be an analogous of the Dirac spheric operator \cite{TH} that reads now  $K=\gamma^0(2 S_3 L_3+\frac{1}{2})$ giving the eigenvalues problems  $K U_{k,\lambda}^{\pm} =\pm\lambda U_{k,\lambda}^{\pm}$. The conclusion is that the fundamental solutions we derived above are eigenspinors of the complete set of commuting operators $\{H,K,J_3, P_3\}$.

The  spinors describing physical states are square integrable packets of a given total angular momentum $\lambda$ having the form
\begin{equation}\label{psilam}
\psi_{\lambda}=\int_{-\infty}^{\infty}dk \left[a_+(k) U^+_{k,\lambda}+a_-(k) U^-_{k,\lambda}\right]
\end{equation} 
where the functions $a_{\pm}$ satisfy the  condition
\begin{equation}\label{Calpha}
\int_{-\infty}^{\infty}dk \left[|a_+(k)|^2+|a_-(k)|^2\right]=1
\end{equation}
that assures the normalization condition $\langle \psi_{\lambda}, \psi_{\lambda}\rangle=1$. We say that the packets (\ref{psilam}) describe the states $(\lambda,a)$ in which the expectation value  of the total angular momentum reads $\langle \psi_{\lambda}, J_3 \psi_{\lambda}\rangle= \lambda$ while the polarization degree can be defined as
\begin{equation}
{\cal P}=\langle \psi_{\lambda}, K\psi_{\lambda}\rangle=\lambda\int_{-\infty}^{\infty}dk \left[|a_+(k)|^2-|a_-(k)|^2\right]\,.
\end{equation}

Now we can derive the currents of the fermions in the states $(\lambda,a)$  by using the components of the current density $j_{\lambda}^{\mu}=\overline{\psi}_{\lambda}\gamma^{\mu}\psi_{\lambda}$.  We consider first the total circular current 
\begin{equation}
I^c_{\lambda}=R\int_{-\infty}^{\infty} dz \overline{\psi}_{\lambda}\gamma^{\phi}\psi_{\lambda}\,,
\end{equation} 
that can be calculated according to Eqs. (\ref{gamph}), (\ref{psiUp}) and (\ref{psiUm}), that yield 
\begin{equation}
\overline{u}^{\pm}_{\lambda}\gamma^{\phi}{u}^{\pm}_{\lambda}=\frac{\lambda+\beta}{2\pi R^3 E_{k,\lambda}}\,, \quad \overline{u}^{\pm}_{\lambda}\gamma^{\phi}{u}^{\mp}_{\lambda}=0\,.
\end{equation}
Then, observing that the integral over the $z$ axis generates a $\delta$-function, we obtain  the definitive closed form
\begin{equation}
I_{\lambda}^c=\frac{\lambda+\beta}{2\pi R^2}\int_{-\infty}^{\infty}\frac{dk}{E_{k,\lambda}} \left[|a_+(k)|^2+|a_-(k)|^2\right]\,.
\end{equation}
Hereby we draw the conclusion  that the circular current is stationary (i. e. independent on $t$)
depending only on the packet content as given by the arbitrary functions $a_{\pm}$. 

Nevertheless, it is remarkable that there are two important properties of the circular currents that are independent on the form of  these functions. The first one is the saturation effect for increasing total angular momenta,
\begin{equation}
\lim_{\lambda \to \pm \infty}I^c_{\lambda}=\pm \frac{1}{2\pi R}\int_{-\infty}^{\infty}dk \left[|a_+(k)|^2+|a_-(k)|^2\right] =\pm \frac{1}{2\pi R}\,.
\end{equation}  
On the other hand,  bearing in mind that the expectation value of the energy in the state $(\lambda,a)$  reads
\begin{equation}
E_{\lambda}=\int_{-\infty}^{\infty}{dk}{E_{k,\lambda}} \left[|a_+(k)|^2+|a_-(k)|^2\right]\,,
\end{equation}
we recover the familiar formula
\begin{equation}
I^c_{\lambda}=\frac{1}{2\pi}\frac{\partial E_{\lambda}}{\partial\beta}\,,
\end{equation}
that has the same form as in the case of the relativistic  \cite{CBC} or non-relativistic AB rings.

Note that, in contrast to the circular current, the longitudinal one,
\begin{equation}
I^3_{\lambda}=R\int_{0}^{2\pi} d\phi\, \overline{\psi}_{\lambda}\gamma^{3}\psi_{\lambda}\,,
\end{equation} 
depends on time, reflecting thus the propagation and dispersion of the packet along the $z$ axis. The general expression of this current  is presented in the Appendix A.

\section{Currents on finite AB cylinders}

Another interesting problem is of a finite cylinder of length $L$ for which we must consider the boundary conditions $f_{1,2}(0)=f_{1,2}(L)=0$. Therefore, we may chose
\begin{equation}\label{fkn1}
\left(
\begin{array}{c}
f_1(z)\\
f_2(z)
\end{array}\right)= N \sin( k_nz)\,\xi_{\sigma} \,,\quad k_n=\frac{\pi n}{L}\,,n=1,2,...\,,
\end{equation}
denoting now $E_{n,\lambda}=E_{k_n,\lambda}$. Thus we obtain the countable discrete energy spectrum 
\begin{equation}\label{Enm}
E_{n,\lambda}=\left[ M^2+\frac{\pi^2 n^2}{L^2}+\frac{1}{R^2}\left(\lambda+\beta\right)^2  \right]^{\frac{1}{2}}\,,
\end{equation}
corresponding to the  square integrable spinors whose components are given by Eq. (\ref{fkn1}) and Eq. (\ref{g12}) that yields now
\begin{equation}
\left(
\begin{array}{c}
g_1(z)\\
g_2(z)
\end{array}\right)=\frac{i N}{E+M}\left(
\begin{array}{cc}
- k_n\cos(k_n z)& -\frac{1}{R}(\lambda+\beta)\sin(k_n z)\\
\frac{1}{R}(\lambda+\beta)\sin(k_n z)& k_n\cos(k_n z)
\end{array}\right) \xi_{\sigma}\,.
\end{equation}
Then, according to Eq. (\ref{psi}) we can write down the form of two types of solutions corresponding to $\sigma =\pm\frac{1}{2}$. For $\sigma =\frac{1}{2}$ we obtain
\begin{equation}\label{psiUp}
U_{n,\lambda}^+(t, \phi,z)=N_{n,\lambda}^+\left(
\begin{array}{r}
\sin(k_n z)e^{i\phi(\lambda-\frac{1}{2})}\\
0\hspace*{20mm}\\
\frac{-ik_n}{E_{n,\lambda}+M}\cos(k_n z)e^{i\phi(\lambda-\frac{1}{2})}\\
\frac{i(\lambda+\beta)}{R(E_{n,\lambda}+M)}\sin(k_n z)e^{i\phi(\lambda+\frac{1}{2})}
\end{array}\right) e^{-iE_{n,\lambda}t}\,,
\end{equation}
and similarly for $\sigma=-\frac{1}{2}$,
\begin{equation}\label{psiUm}
U_{n,\lambda}^-(t, \phi,z)=N_{n,\lambda}^-\left(
\begin{array}{r}
0\hspace*{20mm}\\
\sin(k_n z)e^{i\phi(\lambda+\frac{1}{2})}\\
\frac{-i(\lambda+\beta)}{R(E_{n,\lambda}+M)}\sin(k_n z)e^{i\phi(\lambda-\frac{1}{2})}\\
\frac{ik_n}{E_{n,\lambda}+M}\cos(k_n z)e^{i\phi(\lambda+\frac{1}{2})}
\end{array}\right) e^{-iE_{n,\lambda}t}\,.
\end{equation}
After a little calculation we find that by fixing the value of the normalization constants as
\begin{equation}
N_{n,\lambda}^{\pm}=\frac{1}{\sqrt{\pi RL}}\sqrt{\frac{E_{n,\lambda}+M}{2E_{n,\lambda}}}\,,
\end{equation}
we obtain the desired orthogonality relations
\begin{equation}
\langle U^{\pm}_{n,\lambda}, U^{\pm}_{n',\lambda'}\rangle=\delta_{n,n'}\delta_{\lambda,\lambda'}\,,\quad \langle U^{\pm}_{n,\lambda}, U^{\mp}_{n',\lambda'}\rangle=0\,.
\end{equation}
 The conclusion is that the above fundamental solutions are eigenspinors of the set of commuting operators $\{H,K,J_3, P_3^2\}$.

Let us consider the fermions in the states $(n,\lambda)$ given by the normalized linear combinations  
\begin{equation}\label{psicc}
\psi_{n,\lambda}=c_+ U_{n,\lambda}^+ +c_-U_{n,\lambda}^-\,, \quad |c_+|^2+|c_-|^2=1\,,
\end{equation}
which satisfy $\langle \psi_{n,\lambda}, \psi_{n',\lambda'}\rangle=\delta_{n,n'}\delta_{\lambda,\lambda'} $. The constants $c_{\pm}$ give the polarization degree defined as in the previous case, 
\begin{equation}\label{pol}
{\cal P}=\langle \psi_{n,\lambda}, K \psi_{n,\lambda}\rangle=\lambda (|c_+|^2-|c_-|^2)\,.
\end{equation}
Obviously,  the fermions are unpolarized when $|c_+|=|c_-|=\frac{1}{\sqrt{2}}$.

With these ingredients we can calculate the quantities
\begin{eqnarray}
\overline{U}^{\pm}_{n,\lambda}\gamma^{\phi}{U}^{\pm}_{n,\lambda}&=&\frac{\lambda+\beta}{2\pi R^3 L E_{n,\lambda}}\sin^2 k_n z\,,\\ \overline{U}^{\pm}_{n,\lambda}\gamma^{\phi}{U}^{\mp}_{n,\lambda}&=&\frac{2n}{RL^2E_{n,\lambda}}\sin k_n z\cos k_n z\,.\label{usles}
\end{eqnarray}
that help us to derive the definitive form of the circular currents in the states $(n,\lambda)$ that read
\begin{equation}\label{Inm}
I^c_{n,\lambda}=R\int_{0}^{L} dz\, \overline{\psi}_{n,\lambda}\gamma^{\phi}\psi_{n,\lambda}=\frac{\lambda+\beta}{2\pi R^2 E_{n,\lambda}}=\frac{1}{2\pi}\frac{\partial E_{n,\lambda}}{\partial\beta}\,,
\end{equation}
since the integral over $z$ vanishes the mixed terms (\ref{usles}) while the constants $c_{\pm}$ satisfy Eq. (\ref{psicc}). It is remarkable that this current is independent on polarization and  has a similar form and relation with the energy as in the case of the AB rings  \cite{CBC}. The difference is that now the energy (\ref{Enm}) depends on two quantum numbers, $n$ and $\lambda$, as well as on the length $L$ of the AB cylinder. 

The longitudinal current vanishes since we used boundary conditions that guarantee that
\begin{equation}
\overline{U}^{\pm}_{n,\lambda}\gamma^{3}{U}^{\pm}_{n,\lambda}=\overline{U}^{\pm}_{n,\lambda}\gamma^{3}{U}^{\mp}_{n,\lambda}=0\,.
\end{equation}
In other words, our boundary conditions are of the MIT type vanishing the currents but without canceling  all the components of the Dirac spinors on boundaries.  

\section{Persistent currents on finite AB cylinders}

The properties of the currents (\ref{Inm}) can be better understood by introducing the appropriate dimensionless parameters 
\begin{equation}
\mu=MR\,,\quad \nu=\frac{\pi R}{L}\,, 
\end{equation} 
that allow us to write
\begin{equation}\label{Ichi}
I^c_{n,\lambda}=\frac{1}{2\pi R}\chi_{\mu,\nu}(n,\lambda)\,, \quad \chi_{\mu,\nu}(n,\lambda)=\frac{\beta+\lambda}{\sqrt{\mu^2+\nu^2 n^2+(\beta+\lambda)^2}}\,,
\end{equation}
pointing out the function $\chi$ which gives the behavior of the circular currents.
This function is smooths with respect to all of its variables increasing monotonously with $\lambda$ from $-1$ to $1$ since 
\begin{equation}
\lim_{\lambda\to \pm\infty}\chi_{\mu,\nu}(n,\lambda)=\pm 1\,,
\end{equation} 
and vanishing for $n\to \infty$. This means that, as in previous case, for increasing total angular momenta, the circular current tends to the asymptotic saturation values $\pm (2\pi R)^{-1}$ just  as it happens with the partial currents in AB rings \cite{CBC}.

Now we can use these properties for estimating the persistent current at $T=0$ in semiconductor AB cylinders where the electron discrete energy levels $E_{n,\lambda}$  are given by Eq (\ref{Enm}).  According to the Fermi-Dirac statistics, at $T=0$   
the electrons occupy  all the states $(n,\lambda)$  which satisfy the condition
\begin{equation}\label{Cond0}
E_{n,\lambda}\le E_F+M
\end{equation}
where $E_F\ll M$ is the (non-relativistic) energy of the Fermi level. Therefore, the total number of electrons $N_e$ and the persistent current $I$ can be calculated as
\begin{eqnarray}
N_e&=&\sum_{n,\lambda; E_{n,\lambda}\le E_F+M} 1=\sum_{n,\lambda>0; E_{n,\lambda}\le E_F+M} 2\,,\label{Ne}\\
I&=&\sum_{n,\lambda; E_{n,\lambda}\le E_F+M}I^c_{n,\lambda}=\sum_{n,\lambda>0; E_{n,\lambda}\le E_F+M} (I^c_{n,\lambda}+I^c_{n,-\lambda})\,.
\end{eqnarray}
In practice the flux parameter $\beta$ remains very small (less than $10^{-8}$) such that we can neglect the terms of the order $O(\beta^2)$ of the Taylor expansions of our functions (\ref{Ichi}).  Thus we can write 
\begin{eqnarray}
2\pi R(I^c_{n,\lambda}+I^c_{n,-\lambda})&=&\chi_{\mu,\nu}(n,\lambda)+\chi_{\mu,\nu}(n,-\lambda)\nonumber\\
&=&2j_{\mu,\nu}(n,\lambda) \beta +O(\beta^3)\,,
\end{eqnarray}
where
\begin{equation}
j_{\mu,\nu}(n,\lambda)=\frac{\mu^2+\nu^2 n^2}{(\mu^2+\nu^2 n^2+\lambda^2)^{\frac{3}{2}}}\,,
\end{equation}
obtaining thus the expression of the relativistic persistent currents,  
\begin{equation}\label{Ic}
I=\frac{\beta}{\pi R}\,c(\mu,\nu) \,, \quad c(\mu,\nu)=\sum_{n,\lambda>0;
 E_{n,\lambda}\le E_F+M}j_{\mu,\nu}(n,\lambda)\,.
\end{equation}

The principal problem in evaluating such sums is the computation of the contributing states $(n,\lambda)$ (with $n=1,2,...$ and $\lambda=\pm\frac{1}{2},\pm\frac{3}{2},...$) which  satisfy the condition (\ref{Cond0}).   We denote first by $n_F$ the greatest value of the quantum number $n$ and by $\lambda_n$ the greatest value of $|\lambda|$ for a given $n$, assuming that the states $(n_F,\pm\frac{1}{2})$ and respectively $(n,\pm\lambda_n)$ (with $n=1,2,...,n_F) $ are very close to the Fermi level, i. e. $E_{n_F,\pm\frac{1}{2}}\simeq E_{n,\pm\lambda_n}\simeq E_F+M$. In addition, we denote by $\lambda_F=\lambda_{n=1}$ the greatest value among the quantities $\lambda_n$. Then we can rewrite Eq. (\ref{Cond0}) as
\begin{equation}\label{Cond}
\nu^2 n^2+\lambda^2\le \alpha^2\,, \quad \alpha=R\sqrt{E_F(E_F+2M)}\simeq R\sqrt{2M E_F}\,,
\end{equation}
obtaining the approximative identities
\begin{equation}\label{ident}
\nu^2 {n_F}^2+\frac{1}{4}\simeq \nu^2 n^2 + {\lambda_n}^2\simeq \nu^2+{\lambda_F}^2\simeq \alpha^2
\end{equation}
that help us to estimate the numbers $n_F$ and $\lambda_n$.  Moreover, we observe that  $\lambda_F$  must be much greater than $1$ since otherwise we cannot speak about statistics. Then  we can use the approximative formula  (\ref{Bsum}) obtaining the compact results
\begin{eqnarray}
N_e&=&\sum_{n=1}^{n_F}\sum_{\lambda=\frac{1}{2}}^{\lambda_n}2=
\sum_{n=1}^{n_F}(2\lambda_n+1)={n_F}+2\sum_{n=1}^{n_F} \lambda_n\,, \\    c(\mu,\nu)&=&\sum_{n=1}^{n_F}\sum_{\lambda=\frac{1}{2}}^{\lambda_n}j_{\mu,\nu}( n,\lambda)\simeq\frac{1}{\sqrt{\mu^2+\alpha^2}}\sum_{n=1}^{n_F} \lambda_n\,,
\end{eqnarray}
that represent a very good approximation for the systems with $\mu>200$ \cite{CBC}. It remains to calculate on computer the sum over $n$ or to consider the estimation (\ref{Bsum1}) when $n_F\gg 1$.

An interesting case is of the very short cylinders with $1\ll\nu <\alpha<2\nu$ whose quantum number $n$ can take the unique value $n=n_F=1$ in order to satisfy Eq. (\ref{Cond}) that  becomes now $\lambda^2\le \alpha^2-\nu^2=\lambda_F^2$.  Consequently, the allowed states are $(1,\pm\frac{1}{2}), (1,\pm\frac{3}{2}),...(1,\pm\lambda_F)$ which means that $N_e=2 \lambda_F+1$ and
\begin{equation}\label{Ishort}
I_{short}\simeq \frac{\beta}{\pi R}\frac{\lambda_F}{\sqrt{\mu^2+\alpha^2}}= \frac{\beta}{\pi R}\sqrt{\frac{\alpha^2-\nu^2}{\alpha^2+\mu^2}}\,. 
\end{equation}
However,  for the very short cylinders with $\nu>\alpha$ the identities (\ref{ident}) do not make sense such that we need to rebuild  the entire theory without  motion along the $z$ axis ($k=0$), retrieving thus the case of the ideal AB rings \cite{CBC} for which we must substitute $\nu=0$ and $\alpha=\lambda_F$ in Eq. (\ref{Ishort}). Finally, we note that for the non-relativistic short AB cylinders with $\alpha\ll \mu$ we recover the well-known result, 
\begin{equation}
I_{nr}\simeq\frac{\beta}{\pi R}\frac{\lambda_F}{\mu}\simeq\frac{\beta}{\pi R}\frac{N_e}{2\mu}\,,
\end{equation}
of the non-relativistic persistent current in AB rings.

\section{Concluding remarks}

We presented the relativistic theory of the Dirac fermions on AB cylinders based on the complete $(1+3)$-dimensional Dirac equation with restricted orbital degrees of freedom but without affecting the spin ones. This can be achieved by using the method we proposed recently 
for the AB rings \cite{CBC} according to which the orbital restrictions must be imposes on the  Lagrangian density giving rise to a correct self-adjoint Dirac operator.

The results obtained here point out two principal features of the circular currents on the AB cylinders. The first one is the relation between the circular current and the derivative of the energy with respect to the flux parameter that is the same for AB rings or cylinders either in  the relativistic approach or in the non-relativistic one. In other words this property is universal for all the AB systems with cylindric symmetry. The second feature is specific only for the relativistic circular currents that tend to saturation in the limit of very large total angular momenta, in contrast with the non-relativistic ones that are increasing linearly to infinity. 

The saturation effect determines a specific form of the relativistic persistent current on finite AB cylinders  that can be seen as a generalization of the persistent current in AB rings. Obviously, the dependence on parameters is more complicated in the case of the finite cylinders but these models are closer to the real devices involved in experiments. On the other hand, we must specify that the closed formulas derived in section 5 are only approximations that must be used prudently and completed by numerical calculations on computers.   

We believe that only in this manner,  by using combined analytical and numerical methods, we could step the threshold to the relativistic physics of the Aharonov-Bohm systems.

\appendix

\subsection*{Appendix A: Longitudinal currents}

On the infinite AB cylinders the longitudinal currents in the state $(\lambda,a)$ are determined by the structure of the wave packet (\ref{psilam}) as   
\begin{eqnarray}
I^3_{\lambda}&=&R\int_{0}^{2\pi} d\phi\overline{\psi}_{\lambda}\gamma^{3}\psi_{\lambda}=\frac{1}{4\pi}\int_{-\infty}^{\infty}dkdk' \frac{e^{it(E_{k}-E_{k'})-iz(k-k')}}{\sqrt{E_{k}E_{k'}(E_{k}+M)(E_{k'}+M)}}\nonumber\\
&\times&\left\{\left[kE_{k'}+k'E_{k}+M(E_{k}+E_{k'})\right]\left[a_{+}^*(k)a_{+}(k')+a_{-}^*(k)a_{-}(k')\right]\right. \nonumber\\
&-&\left. \frac{i(\lambda+\beta)}{R}(E_{k}-E_{k'})\left[a_{+}^*(k)a_{-}(k')+a_{-}^*(k)a_{+}(k') \right] \right\}\,,
\end{eqnarray}
where the energies are given by Eq. (\ref{Ekm}) and the arbitrary functions $a_{\pm}$ satisfy the condition (\ref{Calpha}).

\subsection*{Appendix B: Approximating sums by integrals}

Since  $\lambda_F$  must be much greater than $1$ we can use the approximative formula  
\begin{equation}\label{Bsum}
\sum_{\lambda=\frac{1}{2}}^{\lambda_n}j_{\mu,\nu}( n,\lambda)\simeq \int_{0}^{\lambda_n}d\lambda \,j_{\mu,\nu}( n,\lambda)=\frac{\lambda_n}{\sqrt{\mu^2+\nu^2 n^2+\lambda_n^2}}\simeq\frac{\lambda_n}{\sqrt{\mu^2+\alpha^2}}\,,
\end{equation}
that reproduces the numerical results with a satisfactory accuracy for $\mu>200$ \cite{CBC}. Moreover, when $n_F >100$  we can evaluate    
\begin{eqnarray}\label{Bsum1}
\sum_{n=1}^{n_F} \lambda_n&=& \sum_{n=1}^{n_F} \sqrt{\nu^2(n_F^2-n^2)+\frac{1}{4}}\nonumber\\
&\simeq& \int_{x=0}^{n_F}dx \sqrt{\nu^2(n_F^2-x^2)+\frac{1}{4}}\simeq \frac{1}{4}n_F \left(1+\frac{\pi n_F}{\nu}\right) \,.
\end{eqnarray}

\subsection*{Acknowledgments}

I. I. Cot\u aescu  is supported by a grant of the Romanian National Authority for Scientific Research, Programme for research-Space Technology and Advanced Research-STAR, project nr. 72/29.11.2013 between Romanian Space Agency and West University of Timisoara.

D.-M. B\u alt\u a\c teanu is supported by the strategic grant POSDRU /159/1.5/S /137750, Project “Doctoral and Postdoctoral programs support for increased competitiveness in Exact Sciences research”, cofinanced by the European Social Fund within the Sectoral Operational Programme Human Resources Development 2007 - 2013.

\end{document}